%% file: ms.tex
\documentclass{aastex63}
\usepackage{graphicx}
\usepackage{subfigure}
\usepackage{multirow}
\usepackage{comment}
\usepackage{natbib}
\usepackage{hyperref}
\usepackage{mathtools}
\usepackage{mathrsfs}
\usepackage{fontenc}
\usepackage{color}
\usepackage{url}
\usepackage{hyperref}
\usepackage{gensymb}
\usepackage{pifont}
\newcommand{\lbol}{$L_{\rm bol}$}
\newcommand{\REdd}{$\lambda_\mathrm{E}$}

\newcommand       \Angstrom     {\,{\rm \AA}}

\newcommand       \mum          {{\rm \mu m}}

\newcommand       \usfr         {{M_\odot}\,\rm yr^{-1}}

\newcommand{\etal}{\textrm{et al.\ }}

\newcommand{\eg}{\textrm{e.g., }}

\newcommand{\sfrir}{SFR$_{\rm IR}$}

\newcommand{\sfrneon}{SFR$_{\rm Ne}$}

\font\sevenrm=cmr7 scaled 1000

\begin{document}

\title{The Infrared Emission and Vigorous Star Formation of Low-redshift Quasars}

\shortauthors{Xie et al.}

\author{Yanxia Xie}
\affil{Kavli Institute for Astronomy and Astrophysics, Peking University, Beijing 100871, China}

\author{Luis C. Ho}
\affil{Kavli Institute for Astronomy and Astrophysics, Peking University, Beijing 100871, China}
\affil{Department of Astronomy, School of Physics, Peking University, Beijing 100871, China}

\author{Ming-Yang Zhuang}
\affil{Kavli Institute for Astronomy and Astrophysics, Peking University, Beijing 100871, China}
\affil{Department of Astronomy, School of Physics, Peking University, Beijing 100871, China}

\author{Jinyi Shangguan}
\affil{Max Planck Institute for Extraterrestrial Physics, Garching, Germany}

\begin{abstract} 
The star formation activity of the host galaxies of active galactic nuclei (AGNs) provides valuable insights into the complex interconnections between black hole growth and galaxy evolution.  A major obstacle arises from the difficulty of estimating accurate star formation rates in the presence of a strong AGN.  Analyzing the $1-500$~$\mum$ spectral energy distributions and high-resolution mid-infrared spectra of low-redshift ($z < 0.5$) Palomar-Green quasars with bolometric luminosity $\sim 10^{44.5}-10^{47.5}\rm\,erg\,s^{-1}$, we find, from comparison with an independent star formation rate indicator based on [Ne~II]~12.81~$\mum$ and [Ne~III]~15.56~$\mum$, that the torus-subtracted, total infrared ($8-1000\, \mum$) emission yields robust star formation rates in the range $\sim 1-250\,M_\odot\,{\rm yr^{-1}}$.  Combined with available stellar mass estimates, the vast majority ($\sim 75\%-90\%$) of the quasars lie on or above the main sequence of local star-forming galaxies, including a significant fraction ($\sim 50\%-70\%$) that would qualify as starburst systems.  This is further supported by the high star formation efficiencies derived from the gas content inferred from the dust masses.  Inspection of high-resolution Hubble Space Telescope images reveals a wide diversity of morphological types, including a number of starbursting hosts that have not experienced significant recent dynamical perturbations.  The origin of the high star formation efficiency is unknown.
\end{abstract} 


\section{Introduction}\label{sec:intro} 

Supermassive black holes (BHs) are widely regarded as being closely connected with the evolution of galaxies (Richstone \etal 1998; Kormendy \& Ho 2013; Heckman \& Best 2014), but the exact manner in which active galactic nuclei (AGNs) truly impact their host galaxies remains a topic of lively debate.  Does the BH grow in concert with the stars of the host, or does one component lag behind the other?  Much attention has been devoted to the subject of AGN feedback (Fabian 2012), but does AGN feedback inhibit or stimulate star formation?  These issues can be clarified if we have access to the ongoing star formation rate (SFR) and stellar mass ($M_{\rm\ast}$) of the host galaxies of AGNs of different types and in different stages of their evolution.  In the context of the general galaxy population, star-forming galaxies occupy a well-defined main sequence, a relation between SFR and $M_{\rm\ast}$ that encodes vital information on the manner and timescale in which galaxies acquire their stellar mass (\eg Brinchmann \etal 2004; Elbaz \etal 2007; Noeske \etal 2007; Peng \etal 2010; Speagle \etal 2014; Barro \etal 2017).  This same framework serves as a useful guide for probing AGN host galaxies and their relation to the overall galaxy population.

Previous attempts to investigate the star-forming main sequence of AGN host galaxies have yielded mixed results. The global SFRs of low to moderate-luminosity AGNs generally lie on or below the star-forming main sequence for redshifts $0.5 < z < 3$ (\eg Santini \etal 2012; Rosario \etal 2013a; Suh \etal 2017; Jackson \etal 2020), and for the lowest redshifts, low-luminosity AGNs mainly occupy the green valley (\eg Cano-D{\'\i}az \etal 2016; Ellison \etal 2016; Leslie \etal 2016; Catal{\'a}n-Torrecilla \etal 2017) or are effectively passive (Ho et al. 2003).  By contrast, the situation is more controversial for more luminous AGNs and quasars, either locally or at high redshifts. While many studies indicate that luminous AGNs are hosted by normal star-forming galaxies (\eg Harrison \etal 2012; Rosario \etal 2013b; Husemann \etal 2014; Zhang \etal 2016; Stanley \etal 2017; Bernhard \etal 2019; Schulze \etal 2019; Grimmett \etal 2020), others find that they reside above the main sequence (\eg Silverman \etal 2009; Rovilos \etal 2012; Florez \etal 2020; Jarvis \etal 2020; Shangguan \etal 2020b; D. Zhao \etal 2021), some even in the regime of extreme starbursts (\eg  Kirkpatrick \etal 2020). At the same time, seemingly contradictory conclusions are reached by works that report that star formation activity is suppressed in luminous AGNs (\eg Scholtz \etal 2018; Stemo \etal 2020). The root causes of these diverse and potentially conflicting results stem from several factors, including systematics of sample selection, the accuracy of SFR and $M_*$ measurements, and even the very definition of the ``starburst'' phenomenon or the ``main sequence'' now commonly used as a reference to discuss the level of star formation in galaxies.

The cold gas content of AGN host galaxies can shed additional light on these issues.  The increasing availability of cold gas measurements from direct observations of CO and H~I (\eg Evans \etal 2001, 2006; Scoville \etal 2003; Ho \etal 2008a; Wang \etal 2013, 2016; Walter \etal 2014; Xia \etal 2014; Brusa \etal 2015; Husemann \etal 2017; Kakkad \etal 2017; Shangguan \etal 2020a) or from indirect estimates based on dust emission and absorption (\eg Shangguan \etal 2018; Shangguan \& Ho 2019; Yesuf \& Ho 2019, 2020a, 2020b; Zhuang \& Ho 2020) offers the possibility of probing the star formation efficiency (${\rm SFE} \equiv {\rm SFR}/M_{\rm gas}$), which provides a complementary and independent probe of how effectively the cold gas converts to stars in AGN host galaxies (Husemann et al. 2017; S\'anchez \etal 2018; Jarvis et al. 2020; Shangguan et al. 2020b; Zhuang et al. 2021).

As the SFR plays a pivotal role in these considerations, the reliability with which it can be ascertained in AGN host galaxies becomes a central concern.  Measuring the SFR in active galaxies is challenging because essentially all the conventional tracers of young stars in inactive galaxies (Kennicutt 1998a) are subject to contamination, to one extent or another, by radiation from the AGN. The problem is most severe for type~1 (unobscured, broad-line) sources, but type~2 (obscured, narrow-line) systems also suffer. SFRs derived from conventional tracers such as H$\alpha$ are problematic because the AGN narrow-line region itself is a prodigious source of hydrogen recombination emission. Moreover, the mid-infrared (IR) PAH features may be destroyed in the harsh environment of AGNs (Voit 1992). Motivated by Ho \& Keto (2007), Zhuang \etal (2019) proposed an effective SFR estimator for AGNs based on the fine-structure lines of [Ne~II]~12.81~$\mum$ and [Ne~III]~15.56~$\mum$.  The principal limitation of this method is the current scarcity of the requisite mid-IR spectroscopy.  In practice, the optical [O~II]\,$\lambda3727$ line, whose feasibility as a SFR indicator in AGNs was first proposed by Ho (2005; see also Kim et al. 2006) and has been reexamined recently by Zhuang \& Ho (2019), provides a much more readily available alternative, the principal shortcoming being that dust extinction can be a complication (Zhuang \& Ho 2020).  In terms of continuum emission, none of the commonly used SFR indicators from the radio to the ultraviolet can be trusted when the AGN turns on.  The most promising remaining option, albeit not without reservation, lies in the thermal far-IR continuum emission from cool dust grains.  While many contend that the far-IR emission in AGNs primarily traces star formation from the host galaxy (\eg Haas \etal 2003; Schweitzer \etal 2006; Netzer \etal 2007; Lutz \etal 2008; Hatziminaoglou \etal 2010; Santini \etal 2012; Dai \etal 2018), others worry that the AGN can heat dust grains far beyond the limited confines of the small-scale torus (\eg Sanders \etal 1989; Mullaney \etal 2011; Symeonidis \etal 2016; Shimizu \etal 2017; Shangguan et al. 2018; Shangguan \& Ho 2019).   The AGN narrow-line region, after all, contains dust (\eg Groves \etal 2006), and its physical extent can be substantial (Greene et al. 2011; Chen et al. 2019).

This paper offers a fresh appraisal of how well far-IR emission can trace star formation in powerful AGNs, utilizing independent SFRs obtained from the mid-IR neon lines based on the SFR calibration of Zhuang et al. (2019).  We conclude that, at least in low-redshift quasars, far-IR emission, after properly accounting for contamination from the torus, indeed yields robust SFRs.  In concert with estimates of stellar masses and gas masses, we study the star formation properties of the host galaxies and discuss implications for AGN feedback.  Section~\ref{sec:sample_data} presents the data and measurements. In Section~\ref{sec:result}, we compare different SFR tracers, examine the SFRs of quasar hosts relative to the main sequence, and consider their SFEs.  Section~\ref{sec:discussion} discusses the main implications, with a summary given in Section~\ref{sec:conclusion}. We adopt the cosmological parameters $\Omega_m = 0.308$, $\Omega_\Lambda = 0.692$, and $H_{0}=67.8$ km~s$^{-1}$~Mpc$^{-1}$ (Ade \etal 2016).  The SFRs and stellar masses in this paper are all scaled to the stellar initial mass function of Salpeter (1955).

\section{Data and Measurements \label{sec:sample_data}}

We focus on the sample of 86 $z < 0.5$ quasars\footnote{The Boroson \& Green (1992) sample officially contains 87 PG quasars.  We exclude PG~1226+023 (3C~273) from this study because its IR emission is dominated by synchrotron radiation from a prominent jet, which renders the modeling of the spectral energy distribution (SED) highly uncertain (Shangguan \etal 2018).} from the Palomar-Green (PG) survey (Schmidt \& Green 1983), as summarized in Boroson \& Green (1992).  Originally selected based on ultraviolet/optical colors, PG quasars form a representative sample of luminous, broad-line (type~1) AGNs unbiased with respect to dust or gas content.  The low-redshift subset of PG quasars, in particular, has been extensively and systematically studied, providing a rich legacy of multi-wavelength data and physical parameters for the active nuclei and their host galaxies.  For instance, BH masses can be estimated (Vestergaard \& Peterson 2006; Ho \& Kim 2015) from available optical spectra (Boroson \& Green 1992; Ho \& Kim 2009), which, in combination with optical luminosities or more complete broad-band SEDs (Neugebauer et al. 1987; Sanders et al. 1989; Shang et al. 2011) furnish bolometric luminosities and hence Eddington ratios.  

An important parameter used in our study is the total stellar mass ($M_{\ast}$) of the host galaxy. Zhang \etal (2016) provide estimates of $M_{\ast}$ for 55 of the PG quasars based on analysis of high-resolution Hubble Space Telescope (HST) optical and near-IR images. For the remaining 31 sources, we infer their $M_{\ast}$ from the BH mass ($M_{\mathrm{BH}}$) using the empirical correlation (Greene \etal 2020)\footnote{Equation 1 assumes a ``diet'' Salpeter initial mass function (Bell et al. 2003), which is very similar to Kroupa's (2001) initial mass function.  We adjust the zero point of Equation 1 to be consistent with our assumed  Salpeter initial mass function.}

\begin{equation}
\log \left(\frac{M_{\rm BH}}{M_\odot}\right) = (7.89\pm0.09) + (1.33\pm0.12)\log \left(\frac{M_*}{3\times10^{10}\,M_\odot}\right),
\label{eq:mstar}
\end{equation}

\noindent
which has an intrinsic scatter of 0.65 dex.  Optical or near-IR morphologies have been published for 37 sources.  Supplemented with new HST observations (Y. Zhao et al. 2021), 71 of the 86 sources now have HST images useful for morphological classification.  The host galaxies of PG quasars exhibit diverse morphologies (Table~\ref{tab:basic}).  Apart from elliptical galaxies and obvious major mergers, a number of sources display a normal or only a mildly perturbed disk.  For the purposes of the present study, we consider all systems with a disturbed morphology or that possess a nearby companion as mergers. 

\begin{figure}%[ht]
\begin{center}
\includegraphics[width=\textwidth]{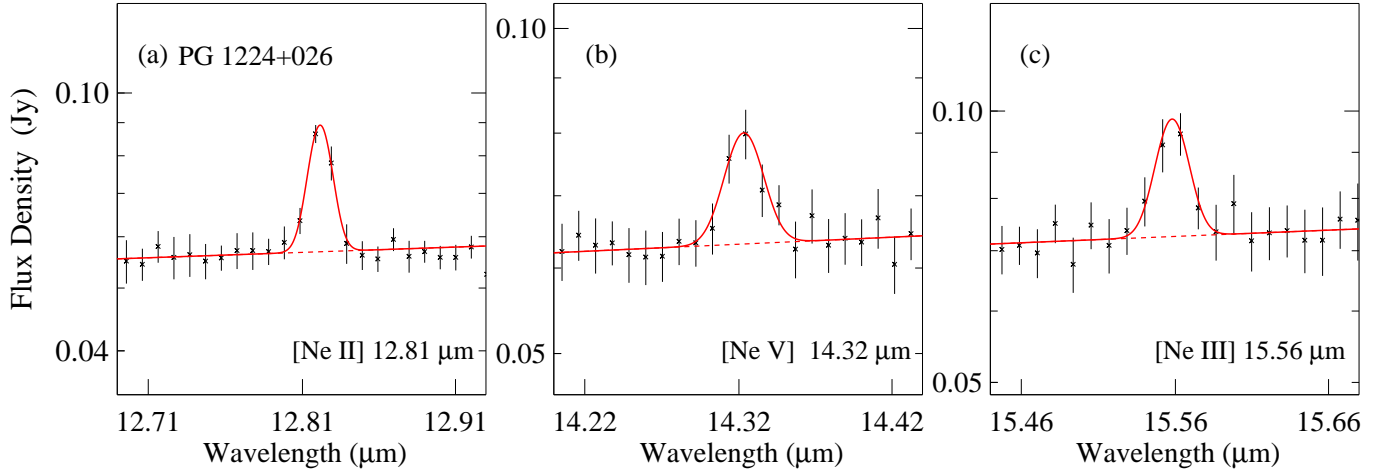}
\caption{Illustration of the fitting of the high-resolution IRS spectrum of PG~1224+026, for the lines (a) [Ne~II] 12.81~$\mum$, (b) [Ne~V] 14.32~$\mum$, and (c) [Ne~III] 15.56~$\mum$. In each panel, the crosses with error bars denote the observed flux density and its corresponding $1\,\sigma$ uncertainty. The red solid curve gives the best-fit single Gaussian profile to the line and the red dashed line marks the underlying continuum.}
\label{fig:fit_neon}
\end{center}
\end{figure}

As our primary aim is to utilize the mid-IR emission lines of [Ne~II]\,12.81~$\mum$, [Ne~III]\,15.56~$\mum$, and [Ne~V]\,14.32~$\mum$ to establish a reliable, independent estimate of the SFR (Zhuang et al. 2019), we rely on high-resolution spectra taken with the Infrared Spectrograph (IRS; Houck et al, 2004) on the Spitzer Space Telescope (Werner et al. 2004), which are available for 37 out of the 86 PG quasars.  The high-resolution spectra comprise data for 34 objects taken with the short-high mode, which samples $9.9 - 19.6\,\mum$ with a $4{\farcs}7 \times 11{\farcs}3$ slit, and data for three higher redshift objects taken with the long-high mode, which covers $18.7 - 37.2\,\mum$ with a $11{\farcs}1 \times 22{\farcs}3$ slit.  Both modes have a full width at half maximum (FWHM) spectral resolution of $\lambda/\Delta\lambda \approx 600$.  We give priority to spectra archived in the Cornell AtlaS of Spitzer/IRS Sources (CASSIS; Lebouteiller \etal 2011, 2015), which extracts spectra according to the spatial extent of each source.  Aperture loss presents a complication for the IRS spectra of our moderately nearby sources, which were obtained with slits of relatively narrow, fixed width.  Fortunately, Shi et al. (2014) performed aperture corrections for the IRS spectra of the PG quasars acquired in the short-low and long-low modes, using MIPS photometry to scale them to the global flux scale. Since the slit widths of the corresponding low-resolution and high-resolution modes are nearly identical, we simply adopt Shi et al.'s aperture corrections for the low-resolution spectra and apply them to the high-resolution spectra.  The aperture correction scaling factors for the vast majority of the sources are smaller than 20\%, which is comparable to the uncertainty of the neon flux measurements.  The three neon lines, being narrow, relatively well-separated from neighboring features and barely resolved in the short-high spectrum, can be measured straightforwardly using a single Gaussian fit on top of a local continuum (Figure~\ref{fig:fit_neon}). The continuum regions are defined on both sides of each line and interpolated. We apply bootstrap resampling to estimate the final line flux and its uncertainty. The significance of the line measurements is evaluated with the quantity $S_{\mathrm {Ne}}$, defined as the ratio of the integrated line flux to the FWHM of the line. We consider a line detected if $S_{\mathrm {Ne}} \geq 3\,\sigma$, with $\sigma$ the standard deviation of the flux density of the local continuum (Xie \& Ho 2019).  We calculate $3\,\sigma$ upper limits by fixing the line width to the observed median value of the detected sources (FWHM = 459 km~s$^{-1}$ for [Ne~II] and FWHM = 582 km~s$^{-1}$ for [Ne~III] and [Ne~V]). A total of 26 sources are detected in [Ne~II], [Ne~III], and [Ne~V]; 11 sources have upper limits on [Ne~V], among them seven with upper limits on [Ne~II]; and one has none of the three lines detected.

From Zhuang \etal (2019),

\begin{equation}
{\rm SFR_{Ne}} \ (M_{\odot}\ {\rm yr^{-1}}) = 4.34\times10^{-41}  \
\left( Z_{\odot}/Z \right) \
\left[ { { L_{\rm [Ne~{\sc II}]+ [Ne~{\sc III}]} - 0.987 { L_{\rm [Ne~{\sc V}]}}}\over{ f_{+} + 1.67 f_{+2} } } \right],
\label{eq:sfr_ne}
\end{equation}

\noindent
where $L_{\rm [Ne~{\sc II}]+ [Ne~{\sc III}]}$ is the luminosity of the sum of [Ne~II]\,12.81~$\mum$ and [Ne~III]\,15.56~$\mum$, $L_{\rm [Ne~{\sc V}]}$ is the luminosity of [Ne~V]\,14.32~$\mum$ (both in units of $\mathrm{erg\ s}^{-1}$), $Z$ is the gas-phase metallicity of the galaxy, and $f_{+}$ and $f_{+2}$ are the fractional abundances of singly and doubly ionized neon, which can be determined from Equation~8 in Zhuang \etal (2019).  We infer the metallicity of the host galaxy indirectly from the stellar mass-metallicity relation (see Section~4.1 of Xie \& Ho 2019). Since most quasar hosts have $M_* \gtrsim 10^{11}\,M_{\odot}$ (Section~\ref{subsec:mse}), above which the stellar mass-metallicity relation flattens (\eg Kewley \& Ellison 2008), we simply adopt the metallicity inferred at $M_*\,=\,10^{11}\,M_{\odot}$ for quasars more massive than this value. We use bootstrapping resampling to estimate the total error budget for $\rm SFR_{Ne}$, taking into account uncertainties from the metallicity, line measurement, and other parameters involved in Equation~\ref{eq:sfr_ne}.  The ionization state of an AGN reflects its Eddington ratio, which, because of observational selection effects in most samples, is also related to its absolute luminosity (Ho 2008, 2009).  The neon-based SFR formalism of Zhuang et al. (2019) only applies to high-ionization quasars and classical Seyfert nuclei, AGNs sufficiently powerful to emit detectable [Ne~{\sc V}]~14.32~$\mum$.  Without the presence of [Ne~{\sc V}]~14.32~$\mum$ as a guide, it would be impossible to ascertain the extent to which the lower ionization transitions of [Ne~{\sc II}] and [Ne~{\sc III}] might arise from the narrow-line region.  In low-accretion rate, low-ionization AGNs (e.g., LINERs), most of the neon emission emerges as [Ne~{\sc II}] and [Ne~{\sc III}].  Since the quasars in our sample are incontrovertibly luminous AGNs, the non-detection of [Ne~{\sc V}] in 12 of the sources must be due to observational sensitivity, and we treat the [Ne~{\sc V}] non-detections as upper limits. The upper limits in [Ne~{\sc V}] and in the lower-ionization lines translate to a range of allowed $\rm SFR_{Ne}$, depending on the limiting values of $f_{+}$ and $f_{+2}$ that appear in the denominator of Equation~2.

To evaluate whether reliable SFRs can be derived using the total IR dust continuum emission, we utilize the 70, 100, 160, 250, 350, and 500 $\mum$ photometry acquired by Petric et al. (2015) using the Herschel Space Observatory (Pilbratt et al. 2010).  The scan-mode observations of Herschel ensure that all of the far-IR flux is captured.  Shangguan et al. (2018) analyzed these far-IR observations, in combination with the IRS mid-IR spectra and sky-survey photometry at shorter wavelengths, to construct complete SEDs covering $\sim 1-500\,\mum$.  To derive dust masses for the host galaxies, Shangguan \etal (2018) used several physical components to decompose the SEDs, including a population synthesis model for the stars (Bruzual \& Charlot 2003), hot dust emission for the AGN torus (Nenkova et al. 2008), and cold dust emission for the global interstellar medium of the host galaxy (Draine \& Li 2007).  The total IR luminosity ($L_{\rm IR}$) of the host is calculated by integrating the luminosity of the Draine \& Li component of the best-fit model from 8 to $1000\,\mum$, which converts to (Kennicutt 1998a)
 
\begin{equation}
{\rm SFR_{IR}} \ (M_{\odot}\ {\rm yr^{-1}}) = 4.5 \times 10^{-44} L_{\mathrm{IR}} \ ({\rm erg} \, {\rm s}^{-1}).
\label{eq:sfr_ir}
\end{equation}

\noindent
The Draine \& Li model contains four main parameters: (1) $U_\mathrm{min}$, the intensity of the radiation field of the diffuse interstellar medium, which scales with the dust temperature; (2) $\gamma$, the mass fraction of the dust in the photodissociation region; (3) $q_\mathrm{PAH}$, the mass fraction of the dust in the form of polycyclic aromatic hydrocarbons (PAHs); and (4) $M_d$, the total dust mass, which sets the overall normalization of the model.  The SEDs of 11 objects are significantly influenced by non-detections in the Herschel bands.  Following Shangguan et al. (2018), we adjust the parameters in the fit that would produce the most conservative upper limit on $L_\mathrm{IR}$ consistent with the far-IR upper limits. After much experimentation, we choose to fix $U_\mathrm{min}=1$ and $q_\mathrm{PAH}=0.47$, as they barely affect the final results.  The parameter $\gamma$ directly correlates with $L_\mathrm{IR}$ but only impacts the global fit negligibly because of the dominance of the AGN torus in the mid-IR. We fix $\gamma$ in the range $0.01 - 0.3$, adjusting only $M_d$ to best match the far-IR upper limits.  A value of $\gamma = 0.3$ is consistent with the upper range of values derived from the well-constrained fits in the sample (see Figure~6 of Shangguan et al. 2018).  We present neon line luminosity, IR luminosity, and all SFR measurements in Table~\ref{tab:sfrqso}.

\begin{figure}%[ht]
\begin{center}
\includegraphics[height=0.6\textwidth]{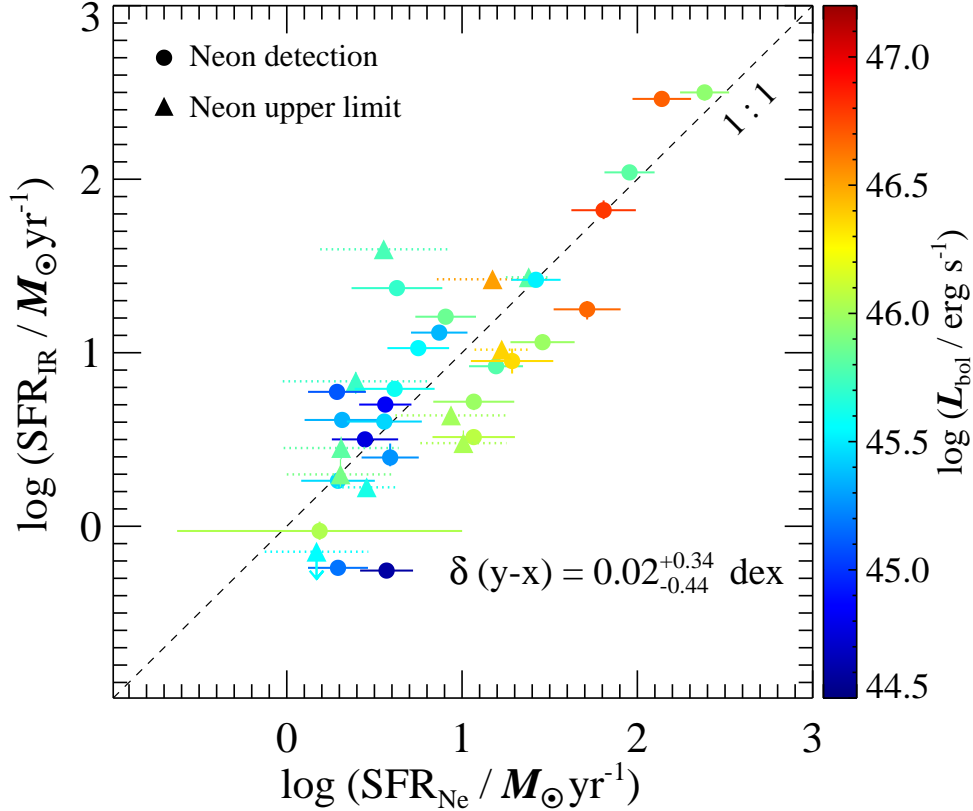}
\caption{Comparison between SFRs derived using the total IR ($8-1000\,\mum$) luminosity with those derived using the luminosity of the mid-IR neon lines. The data points are color-coded according to \lbol. Circles denote objects with all neon lines detected; triangles mark objects for which one or more of the neon lines is an upper limit, with the dotted line indicating the range of SFR bracketed by Equation~2. The dashed line is the 1:1 relation. The median (16\%, 84\%) of the difference between the two SFRs (log~\sfrir$-$log~\sfrneon) is $0.02_{-0.44}^{+0.34}$ dex.}
\label{fig:sfr_fir_neon}
\end{center}
\end{figure}

\section{Results \label{sec:result}}

\subsection{Far-IR Luminosity as a SFR Estimator for Quasars}

Using the mid-IR neon lines as an independent indicator of SFR in AGNs (Zhuang et al. 2019), we are now in a position to assess the extent to which the far-IR luminosity tracks star formation, for the 37 PG quasars for which both tracers can be measured.  Figure~\ref{fig:sfr_fir_neon} shows that \sfrir\ and \sfrneon\ follow a 1:1 relation, with no noticeable systematic dependence on AGN bolometric luminosity.  Using the Weibull distribution as implemented in the {\tt Python} package {\tt lifelines}\footnote{https://lifelines.readthedocs.io/en/latest/} (Davidson-Pilon et al. 2020) to account for the interval of $\rm SFR_{Ne}$ allowed by the upper limits in the neon lines, we find that the median (16\%, 84\%) of the difference between the two SFRs (log~\sfrir$-$log~\sfrneon) is $0.02_{-0.44}^{+0.34}$ dex.  The consistency between these two measures of SFR suggests that the global dust content of the host galaxies is heated predominantly by young stars, and that the torus-subtracted, total IR ($8-1000\,\mum$) luminosity provides an unbiased estimator of the SFR, even for relatively powerful, unobscured quasars, such as those contained in the low-redshift PG sample.  Reliable uncertainties are difficult to estimate, however.  The total uncertainty of \sfrneon\ is at least $\sim\,0.2$\,dex, dominated by uncertainties in the metallicity estimates, aperture corrections, and the line measurements themselves (Section~\ref{sec:sample_data}).  On the other hand, the formal uncertainties for \sfrir\ or $L_{\rm IR}$, mostly $\lesssim 0.05$\,dex, likely are severely underestimated.  As discussed in Shangguan et al. (2018), the SED fits used to derive $L_{\rm IR}$ sample a limited grid of parameters of the Draine \& Li (2007) models.   From comparison of SED fits using different assumptions for the torus component and analysis of mock data (Appendices~C and D in Shangguan \etal 2018), the true uncertainties on $L_{\rm IR}$ are closer to $\sim0.2-0.3$\,dex.  Despite these complications, we note that, over the 3 orders of magnitude in AGN bolometric luminosity (\lbol\ $\approx 10^{44.5}-10^{47.5}\rm\,erg\,s^{-1}$) and 2.5 orders of magnitude in star formation activity (SFR $\approx 1-250\,M_\odot\,{\rm yr^{-1}}$) spanned by our sample, the observed scatter of the correlation in Figure~\ref{fig:sfr_fir_neon} is only $\sim$\,0.4\,dex.  For the rest of the paper, we adopt, as default, \sfrir\ to measure star formation activity, whose final error budget is assumed conservatively to be 0.4 dex. 
 
\begin{figure}%[ht]
\begin{center}
\includegraphics[height=0.68\textwidth]{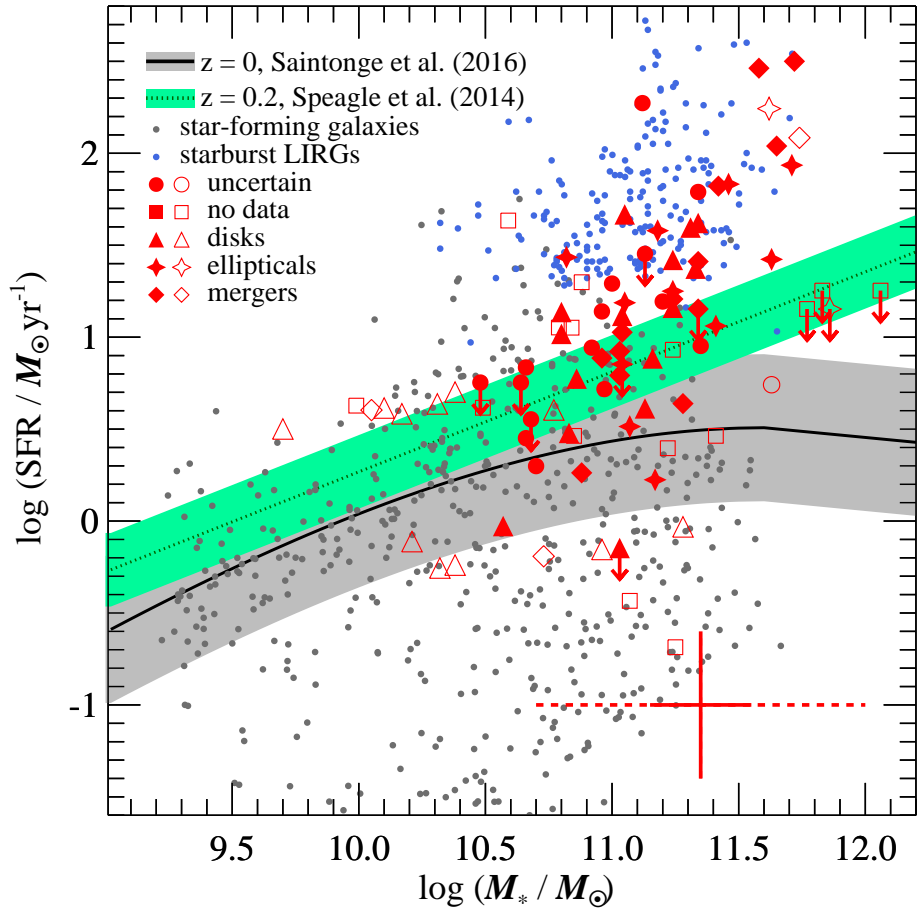}
\caption{{PG quasars lie above the star-forming galaxy main sequence at $z\approx 0$ as defined by Saintonge et al. (2016; black solid line and grey shaded region denoting the $\pm 0.4$ dex $1\,\sigma$ scatter) and at $z\approx 0.2$ as given by Speagle \etal (2014; green dotted line and green shaded region denoting the $\pm 0.2$ dex $1\,\sigma$ scatter).  Filled red symbols indicate objects having direct stellar mass ($M_{\ast}$) measurements from high-resolution optical or near-IR images, while open red symbols represent objects with $M_{\ast}$ estimated from the $M_{\mathrm{BH}} - M_{\ast}$ relation. The typical uncertainties for SFR are 0.4 dex, and for $M_{\ast}$ they are 0.2 dex for direct and 0.65 dex for indirect measurements, as indicated by the red solid and dashed lines, respectively, in the lower-right corner.  Different symbols represent different host galaxy morphologies, as given in the legend.  Shown for reference are star-forming galaxies from the xCOLD GASS sample (Saintonge et al. 2017; filled grey points) and starburst LIRGs (Shangguan et al. 2019; filled blue points).}}
\label{fig:main_sequence}
\end{center}
\end{figure}

\subsection{Quasar Hosts and the Galaxy Star-forming Main Sequence \label{subsec:mse}}

Having established that we can trust the SFRs derived from the torus-subtracted, total IR luminosity, we now make use of \sfrir\ and $M_*$ to examine our sample of quasar hosts in the context of the star-forming galaxy main sequence (Figure~\ref{fig:main_sequence}).  For the present purposes, we adjust all quantities to a common scale normalized to the Salpeter (1955) stellar initial mass function\footnote{Stellar masses and SFRs for the stellar initial mass function of Salpeter (1955), Kroupa (2001), and Chabrier (2003) scale in the ratio 1:1.49:1.58 (Kennicutt \etal 2009; Madau \& Dickinson 2014).}.  For comparison, we show the star-forming galaxies from xCOLD GASS (Saintonge et al. 2017), which comprises low-redshift ($0.01 < z < 0.05$) systems with $M_* > 10^{8}\, M_{\odot}$ whose SFRs were derived from a combination of ultraviolet and mid-IR photometry (Saintonge \etal 2016).  The black curve delineates the parametric relation of the $z \approx 0$ star-forming main sequence from Saintonge \etal (2016), which has a $1\,\sigma$ scatter of $\pm0.4$\ dex as indicated by the gray band.  The majority (69\% or 59/86, including nine upper limits) of the quasar host galaxies formally lie above the main sequence, and thus qualify as ``starbursts.''  Only 19 of the quasar hosts sit on the main sequence, with eight located in the green valley.  Defining the specific SFR as ${\rm sSFR} = {\rm SFR}/M_*$, we use the Kaplan-Meier estimator, as implemented in \texttt{lifelines}, to calculate the median and $16\%-84\%$ interval of $\log ({\rm sSFR}/\rm yr^{-1}) = -10.11^{+0.49}_{-0.78}$ for quasar hosts, much higher than $\log ({\rm sSFR}/\rm yr^{-1}) = -11.17\pm0.73$ for the star-forming galaxies from xCOLD GASS over the mass range of $M_{\ast} \ge 10^{10.5}\,M_{\odot}$ where they mostly overlap.  Considering that the slope, zero point, and possibly even the shape of the star-forming main sequence evolve with cosmic time, perhaps a more appropriate reference should be made to the  main sequence of Speagle \etal (2014), which is defined for star-forming galaxies at $z\,\approx\,0.2$, similar to the average redshift of the PG quasars.  Relative to this reference frame, nearly half (41/86 or 48\%) of the quasars still formally lie above the $1\,\sigma$ scatter of $\pm$\,0.2\,dex of the higher-redshift main sequence.  Lastly, it is illustrative to consider the luminous IR galaxies (LIRGs) from the Great Observatories All-Sky LIRG Survey (GOALS; Armus et al. 2009), most, if not all, of which are well-known starbursts (blue points in Figure~\ref{fig:main_sequence}).  The GOALS galaxies have uniform photometry in the Herschel bands, and their $1-500\,\mum$ SEDs were analyzed by Shangguan et al. (2019) following exactly the same methodology as Shangguan et al. (2018) employed for the PG quasars.  If we simply use as reference the GOALS sample, which has a minimum ${\rm SFR} \approx 20\,\usfr$, then 25\% (22/86) of the PG quasars overlap with the LIRGs.  Thus, by any reasonable measure, a significant fraction of the PG quasars live in starburst galaxies.

It is interesting to note that the host galaxies of the quasars are highly heterogeneous. Among those that have fairly secure classifications, the morphologies of the host galaxies range from ellipticals and mergers to seemingly undisturbed disk systems, with no obvious correlation with location on the ${\rm SFR}-M_*$ plane.  Disk galaxies can be found on or off the main sequence, including the domain securely occupied by starbursts.  As discussed in Section~4, this finding poses a challenge to the notion that major mergers are a necessary ingredient for triggering quasar activity.

\begin{figure}%[ht]
\begin{center}
\includegraphics[height=0.68\textwidth]{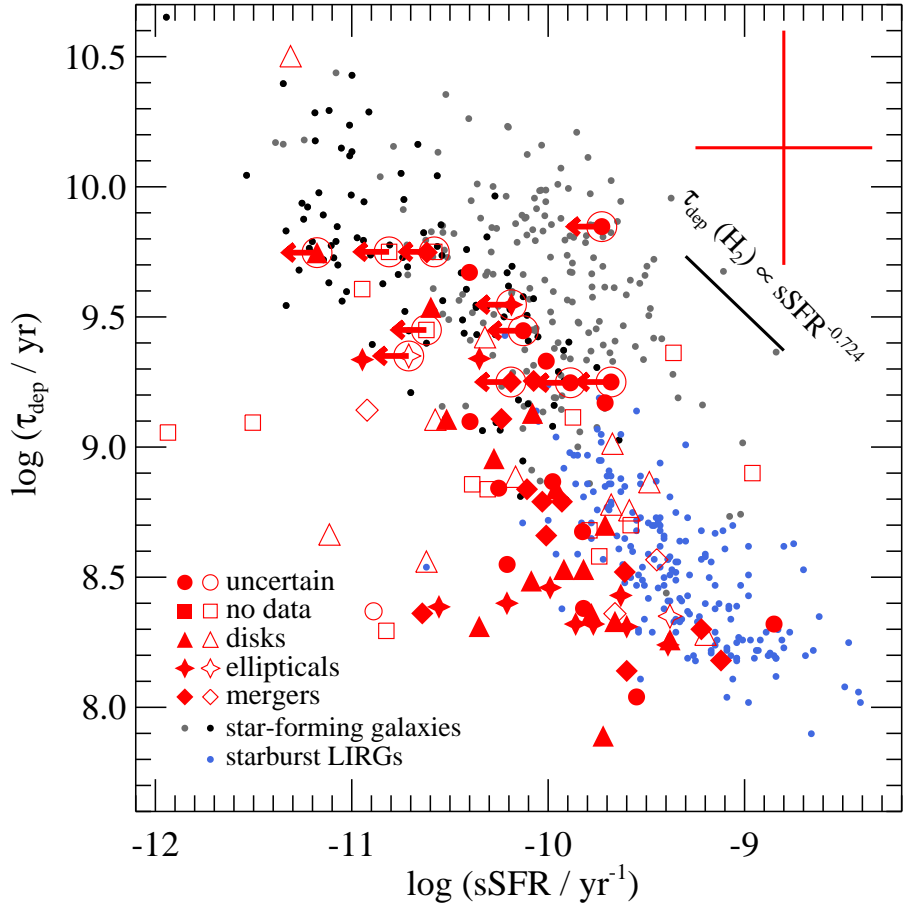}
\caption{{The variation of gas depletion timescale ($\tau_{\rm dep}$) with specific star formation rate (sSFR) for the PG quasars are compared with star-forming galaxies from xCOLD GASS (Saintonge et al. 2017; filled grey points for $M_{\ast} < 10^{10.5}\, M_{\odot}$; filled black points for $M_{\ast} \ge 10^{10.5}\, M_{\odot}$) and starburst LIRGs (Shangguan \etal 2019; filled blue circles). Typical uncertainties for the PG quasars are given in the top-right corner.  Filled red symbols indicate objects having direct $M_{\ast}$ measurements from high-resolution optical or near-IR images, while open red symbols represent objects with indirect $M_{\ast}$ estimated from the $M_{\mathrm{BH}} - M_{\ast}$ relation. Different symbol types represent different host galaxy morphologies, as given in the legend.  Objects that have both gas mass and SFR upper limits are highlighted with a large circle. The solid black line gives the slope of the relation $\tau_{\rm dep} \rm (H_{\sevenrm2}) \propto sSFR^{-0.724}$ for star-forming galaxies with $M_{\ast} \ge 10^{10.5}\, M_{\odot}$ (Saintonge \etal 2011).}}
\label{fig:pg_sfe}
\end{center}
\end{figure}

\subsection{The Global Gas Content and Star Formation Efficiency of Quasar Hosts}\label{subsec:sfe}

The preceding subsection reveals that a significant fraction ($\sim 50\%$) of quasar host galaxies form stars at a rate that places them above the star-forming galaxy main sequence, including $\gtrsim 25\%$ that are situated securely in the regime of starburst galaxies.  Since we have access to total gas mass estimates for the sample (Shangguan et al. 2018), we can study the relation between cold gas content and global star formation activity (Schmidt 1959; Kennicutt 1998b).  Starburst galaxies are characterized by higher SFEs or, equivalently, shorter gas depletion timescales ($\tau_{\rm dep}$ = SFE$^{-1}$) than normal star-forming galaxies (\eg Daddi \etal 2010; Genzel \etal 2010; Shangguan et al. 2019). The majority ($\sim 90$\%) of PG quasars, far from lacking cold gas as might be expected from certain models of AGN feedback, in fact possess similar gas fractions ($M_{\rm gas}/M_* \approx 0.1$) compared to star-forming galaxies. The same holds for low-redshift ($z \lesssim 0.5$) type~2 quasars (Shangguan \& Ho 2019).  As with star-forming galaxies, quasars exhibit a strong inverse correlation between $\tau_{\rm dep}$ and sSFR (Figure~\ref{fig:pg_sfe}). The relation for quasar hosts may be slightly steeper than, but it is not inconsistent with, that for massive ($M_{\ast} \gtrsim 10^{10.5}\, M_{\odot}$) star-forming galaxies defined in terms of molecular gas: $\tau_{\rm dep} ({\rm H_{\sevenrm2}}) \propto$\ sSFR$^{-0.724}$ (Saintonge \etal 2011).  The majority of our sample (64/86 or 74\%) have gas depletion timescales distinctively offset from those of normal galaxies and comparable to those of starburst LIRGs ($\tau_{\rm dep} \,<\,10^{9.2}$\,yr).  The subset of starburst-like quasar hosts covers the entire range of stellar masses and morphological types of the parent sample, including, as mentioned, objects with no obvious evidence of significant ongoing or recent merger activity.  

It is notable that a handful (10) of the elliptical host galaxies have exceptionally short gas depletion timescales (median $\tau_{\rm dep} = 10^{8.4}$\,yr.  All are massive (median $M_{\ast} = 10^{11.3}\,M_{\odot}$), presumably advanced or recent merger remnants.  While their gas fraction is rather modest ($M_{\rm gas}/M_* \approx 0.04$), they have sufficiently healthy levels of star formation (${\rm SFR} \approx 27$ $\usfr$) that would qualify most of them as starbursts in terms of SFE.  The starburst activity in these systems may be related to central gas concentration from recent merger activity (see, e.g., Husemann et al. 2017 for similar arguments for their sample), but high-resolution, high-sensitivity observations are needed to confirm or refute this hypothesis.

\section{Discussions \label{sec:discussion}}

\subsection{The Reliability of Infrared-based SFRs in AGN Host Galaxies}

The heating source of the IR emission of AGNs continues to be a topic of intense debate. In light of the abundant cold interstellar medium detected in the host galaxies of luminous AGNs (\eg Scoville \etal 2003; Ho \etal 2008b; Wang \etal 2013; Walter \etal 2014; Xia \etal 2014; Husemann \etal 2017; Kakkad \etal 2017; Shangguan \etal 2018; Shangguan \& Ho 2019; Li et al. 2020; Zhuang \& Ho 2020), partitioning between the contribution of BH accretion and young stars to the IR luminosity is inherently ambiguous and fraught with uncertainty.  Depending on the spatial distribution of the interstellar medium, dust grains can be exposed to the AGN even on large scales, large enough to reprocess the primary radiation in the far-IR (Sanders et al. 1989).  The existence of extended narrow-line regions (e.g., Greene et al. 2011; Chen et al. 2019) attests to the potential reach of the AGN radiation field.  Thus, no place in the host galaxy can be assumed to be immune from AGN heating.  There are several strategies to confront this problem: (1) remove the likely strength of  the AGN component in the IR by scaling to an empirical estimate of the strength of the AGN anchored in another band (e.g., hard X-rays: Dai et al. 2018) or by modeling the contribution of the AGN-heated torus (e.g., Hatziminaoglou \etal 2010; Zhuang et al. 2018); (2) subtracting an empirical AGN template that extends into the far-IR (e.g., Kirkpatrick et al. 2020; Li et al. 2020); or (3) choose a sufficiently long wavelength where the AGN contribution is assumed to be negligible (e.g., Rosario \etal 2012; Santini \etal 2012).  However, while plausible, none of these approaches is truly foolproof.  Indeed, based on their analysis of the intrinsic far-IR SED of type~1 AGNs, Symeonidis \etal (2016) argue that no region shortward of $\sim 1000\,\mum$ escapes AGN contamination.  While this assertion has been challenged subsequently (Lani et al. 2017; Lyu \& Rieke 2017; Xu et al. 2020), it remains the case that is extremely difficult to establish quantitatively the true extent to which AGN heating affects the far-IR regime.  The AGN component for sure matters in the most powerful, distant quasars (e.g., Lutz et al. 2008; Li et al. 2020).  Concern also has been raised frequently for low-redshift quasars, indeed for the PG sample itself (Sanders et al. 1989; Haas et al. 2003; Symeonidis et al. 2016; Zhuang et al. 2018).  In their detailed analysis of the $1-500\,\mum$ SED of the PG sample, Shangguan \etal (2018) find that the dust temperature (or the minimum radiation field intensity) of the quasar hosts mildly correlates with the quasar luminosity, suggesting that the AGN may heat the interstellar medium on galactic scales.  The same trend holds in a matched sample of $z < 0.5$ type~2 quasars (Shangguan \& Ho 2019).  The problem, of course, becomes even more acute in lower luminosity AGNs (\eg Mullaney \etal 2011; Shimizu \etal 2017).

Using PAH emission as an independent indicator of star formation, a number of investigators have concluded that young stars mainly power the far-IR continuum of nearby quasars (Schweitzer \etal 2006; Netzer \etal 2007; Lutz \etal 2008).  Barthel (2006) reached a similar conclusion, under the assumption that radio continuum emission faithfully traces the SFR in radio-quiet quasars.  In the same spirit, the current study takes advantage of the new SFR indicator for AGNs based on the mid-IR neon lines recently proposed by Zhuang et al. (2019).  We believe that the mid-IR neon lines provide more secure SFRs than PAH emission, which may be prone to AGN destruction (Voit 1992), or the radio continuum, which is unavoidably confused by AGN emission at some level, even in radio-quiet sources.  The SFR calibration of Zhuang et al. (2019) has an estimated accuracy of $\sim$0.2 dex, primarily dominated by uncertainties in metallicity.  Our parent sample is drawn once again from the low-redshift PG quasars (Boroson \& Green 1992), which has been systematically studied by many, including us.  Importantly, we have previously derived torus-subtracted total IR luminosities (Shangguan et al. 2018), which take into account the most updated torus models and utilize the full information contained in the mid-IR ($5 - 40\,\mum$) Spitzer IRS spectra (Shi et al. 2014).  From detailed SED modeling (Shangguan et al. 2018), we have available accurate total dust masses and hence cold gas masses, whose robustness have been verified with ALMA CO observations (Shangguan et al. 2020a, 2020b).  Among the 86 sources in the original $z < 0.5$ sample (see footnote 1), 37 have usable high-resolution IRS spectra, which are necessary to properly resolve the neon lines from the pervasive PAH features.  While the statistics are still somewhat meager, this subsample does span a significant range in bolometric luminosity [$L_{\rm bol} = 10 \lambda L_{\lambda}\,(5100 \Angstrom) \approx 10^{44.5} - 10^{47.5}\rm erg\,s^{-1}$], BH mass ($M_{\rm BH}\approx 10^{6.5}-10^{9.9}\,M_\odot$), and Eddington ratio ($\lambda_{\rm E} = L_{\rm bol}/L_{\rm Edd} \approx 0.02 - 2$).  The neon-based estimator yields ${\rm SFR} \approx 1-250\,M_\odot\,{\rm yr^{-1}}$, which agree surprisingly well ($0.02^{+0.34}_{-0.44}$ dex; Figure~2) with the SFRs inferred from the torus-subtracted total IR ($8-1000\,\mum$) luminosity.  The $\sim$\,0.4\, dex scatter is still significant, but it is primarily dominated by the systematic uncertainties of the two SFR indicators (see Sections~2 and 3.1).  Our conclusion is consistent with that of Shangguan \etal (2020b), who recently found, based on ALMA CO observations of a subset of the present sample, that the AGN-corrected IR luminosity correlates tightly with the molecular gas content, strongly suggesting that the IR emission accurately traces ongoing star formation in the host galaxies.

\subsection{The Mode of Star Formation in Quasars}

The evolution of most star-forming galaxies is regulated by secular processes that situate them on a relatively tight main sequence with a scatter of $\sim0.2 - 0.3$\,dex on the ${\rm SFR} - M_*$ diagram (\eg Brinchmann \etal 2004; Daddi \etal 2007; Speagle \etal 2014).  The exact slope and shape of the main sequence are controversial for galaxies with $M_*\,>\,10^{10}\,M_\odot$ (\eg Renzini \& Peng 2015; Schreiber \etal 2015; Whitaker \etal 2015; Tomczak \etal 2016), with differences attributable to variations in galaxy internal structure, systematics of sample selection, choice of SFR indicator, and methods adopted for statistical analysis of the ${\rm SFR} - M_*$ distribution (\eg Abramson \etal 2014; Mancini \etal 2019; Popesso \etal 2019).  Gas-rich major mergers can trigger starbursts (\eg Sanders \& Mirabel 1996), which manifest themselves as large excursions above the main sequence, although here, too, there is no universal agreement as to the exact criteria that define a starburst (\eg Elbaz \etal 2011; Rodighiero \etal 2011; Bergvall \etal 2016).

What is the dominant mode of star formation for AGNs in general, and for quasars in particular?  The main sequence gives a useful framework for discussing the evolutionary status of AGN host galaxies and their relation to the galaxy population at large.  The existing literature in this field, however, is complicated enormously by the diverse strategies of AGN sample selection, the accuracy of the SFR and $M_*$ tracers, and the myriad choices of main sequence prescription.  While there is almost unanimous agreement that AGNs of low to moderate luminosity ($L_{\rm bol} \lesssim 10^{45}\, \rm erg\,s^{-1}$) at $z \approx 0-3$ lie on or below the main sequence (e.g., Shao \etal 2010; Mullaney \etal 2012, 2015; Santini \etal 2012; Rosario \etal 2013a; Shimizu \etal 2015; Ellison \etal 2016; Leslie \etal 2016; Suh \etal 2017; Bernhard \etal 2019; Grimmett \etal 2020; Jackson \etal 2020), no consensus exists for AGNs with $L_{\rm bol} > 10^{45}\, \rm erg\,s^{-1}$.  The situation is particularly contentious at redshifts higher than $\sim 0.5$, where luminous AGNs have been reported to be above (\eg Rovilos \etal 2012; Florez \etal 2020; Kirkpatrick \etal 2020), on (\eg Harrison \etal 2012; Xu \etal 2015; Stanley \etal 2017; Schulze \etal 2019), and below (\eg Scholtz \etal 2018; Stemo \etal 2020) the main sequence. Fortunately, better convergence of opinion can be found for luminous ($L_{\rm bol} \gtrsim 10^{45}\, \rm erg\,s^{-1}$) AGNs at $z \lesssim 0.5$. Most agree that low-redshift quasars are located largely on and above the main sequence (Husemann \etal 2014; Xu \etal 2015; Zhang \etal 2016; Stanley \etal 2017; Jarvis \etal 2020).  Regardless of redshift, it appears that the magnitude of an AGN's offset from the main sequence correlates positively with its luminosity (Bernhard et al. 2019; Grimmett et al. 2020).  For their large sample of $z\approx 0.3$ type~1 AGNs with uniform SFRs based on extinction-corrected [O~II]\,$\lambda3727$ emission, Zhuang \& Ho (2020) demonstrate that SFR systematically rises with increasing $L_{\rm bol}$ at fixed $M_*$.

Our current work takes advantage of the increasingly comprehensive set of physical parameters available for the PG quasars, spanning \lbol\ $\approx \rm 10^{44.5} - 10^{47.5}\,erg\,s^{-1}$, for which we now have homogeneously measured, newly calibrated SFRs based on the total IR ($8-1000\,\mum$) luminosity (Section~3.1), stellar masses (Section~2), and cold gas content (Shangguan et al. 2018, 2020a).  Originally selected in the optical/ultraviolet, the PG quasars are not biased toward gas-rich systems, and our analysis covers the {\it entire}\ sample of 86 $z < 0.5$ sources (see footnote 1). The detection rate for the dust (gas) content and IR luminosity (SFR) is very high ($\sim 90\%$). The uniqueness of this dataset, in terms of sample size, sample purity, and data quality affords us an opportunity to revisit the star formation properties of low-redshift quasars.  Depending on whether we adopt the work of Saintonge et al. (2016) or Speagle et al. (2014) as reference, $20\%-30\%$ of the quasars are within the $1\sigma$ scatter of the main sequence, while $50\%-70\%$ scatter above and $10\%-25\%$ fall below it. If, instead, we use GOALS (Armus et al. 2009; Shangguan et al. 2019) as a guide, then roughly $25\%$ (22/86) the quasars intermix with the LIRGs. In short, at least $\sim 25\%$ and as many as $\sim 70\%$ of the PG quasars can be deemed to be starbursts according to their sSFRs.  Despite some of the ambiguities inherent in the interpretation of the ${\rm SFR} - M_*$ diagram, fortunately we can circumvent them using the available gas mass measurements, which, when combined with the SFRs, yield SFEs (Section~3.3). We find that more than half of the quasar hosts will consume their gas reservoir in $\tau_{\rm dep} \approx 0.4\,\rm Gyr$ in the median.  These SFEs are comparable to those of LIRGs (Shangguan \etal 2019), which are prototypical low-redshift starbursts.

Our study qualitatively agrees with and statistically reinforces previous results based on smaller samples of nearby quasars.  In their CO and optical integral field spectroscopic survey of 14 $z < 0.2$ Hamburg/ESO quasars, Husemann \etal (2017) show that, apart from two starburst-like objects undergoing a major merger, the host galaxies are essentially indistinguishable from normal star-forming galaxies in terms of their gas fraction and SFE. Jarvis \etal (2020) obtain CO observations of a handful of $z \approx 0.1$ type~2 quasars hosting kpc-scale ionized outflows and jets and conclude that the majority of them possess higher molecular gas fraction and SFE than star-forming galaxies.

\begin{figure}%[ht]
\begin{center}
\includegraphics[height=0.6\textwidth]{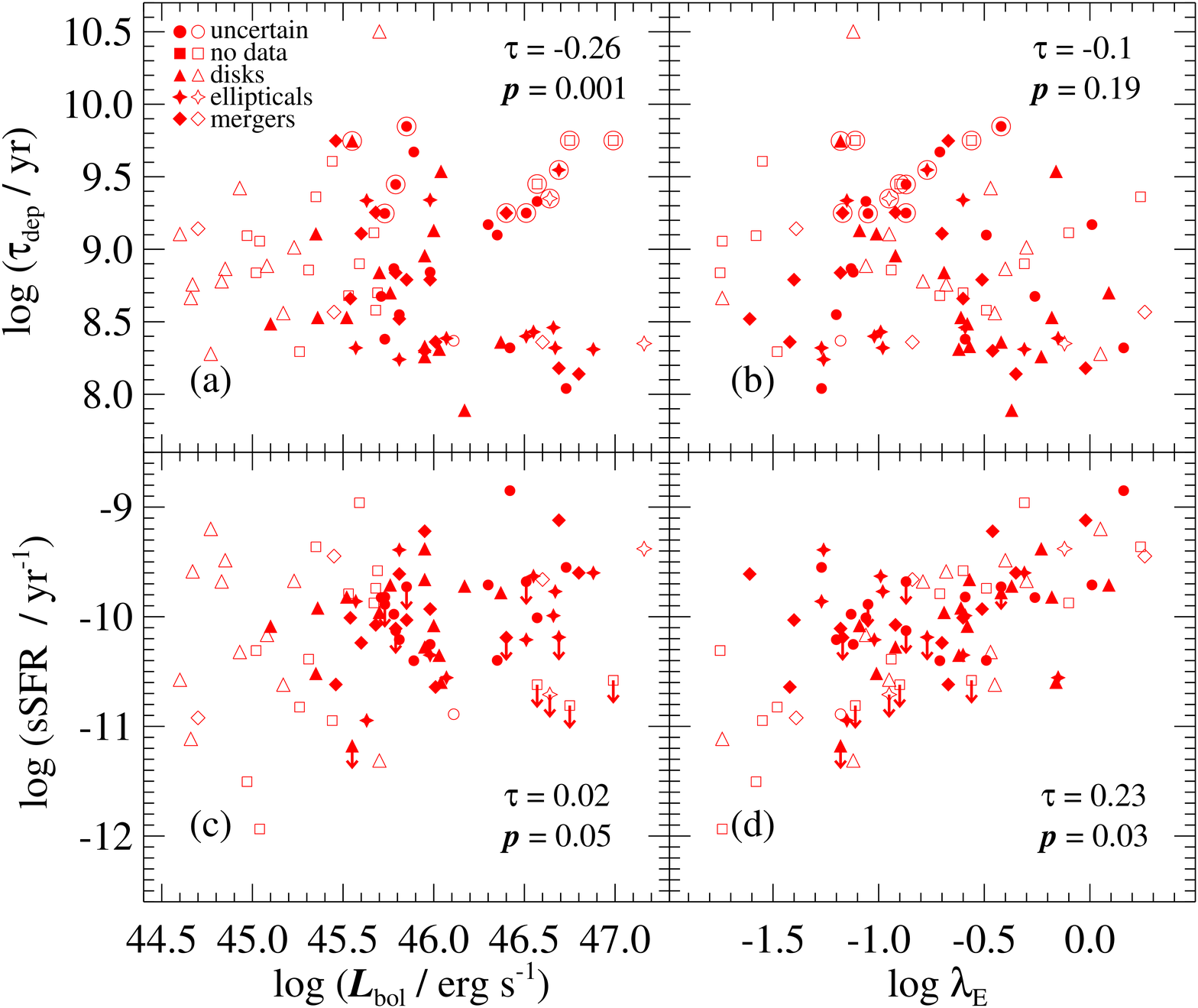}
\caption{Dependence of $\tau_{\rm dep}$ on (a) AGN bolometric luminosity and (b) Eddington ratio. Dependence of sSFR on (c) AGN bolometric luminosity and (d) Eddington ratio. 
Filled red symbols indicate objects having direct $M_{\ast}$ measurements from high-resolution optical or near-IR images, while open red symbols represent objects with indirect $M_{\ast}$ estimated from the $M_{\mathrm{BH}} - M_{\ast}$ relation. In panels (a) and (b), objects that have both gas mass and SFR upper limits are highlighted with a large circle. The Kendall's correlation coefficient $\tau$ and $p$-value derived from the {\tt cenken} function in {\tt R} are given. In the correlation analysis we exclude objects with SFR and gas mass upper limits in panels (a) and (b), and we exclude objects with indirect $M_{\ast}$ in panels (c) and (d).}
\label{fig:ssfr_tdep_agn_par}
\end{center}
\end{figure}

\subsection{The Origin of High SFEs in Quasar Host Galaxies \label{subsec:feedback}}

What triggers starbursts? According to conventional wisdom: gas-rich, major mergers (\eg Kim \etal 2002; Veilleux \etal 2002; Larson \etal 2016).  In this context, the present results for the PG quasars are somewhat puzzling. As discussed in Section~4.2, we have established that the host galaxies of more than half of the quasars have specific SFRs and gas depletion time scales akin to those of starburst systems. Yet, careful scrutiny of Figures~3 and 4 reveals that not all the starbursting hosts display signatures of ongoing mergers or interactions. For example, for the 37 sources with $\tau_{\rm dep} \lesssim 0.5\,\rm Gyr$, 28 have sufficient HST imaging to yield reliable morphologies, and among those only $\sim 29\%$ (8/28) have clear merger signatures. The rest are ellipticals ($\sim 32\%$, 9/28) and---most surprisingly---undisturbed disk galaxies ($\sim 39\%$, 11/28).  The significance of this result is unclear.  Apart from small-number statistics, we note that while starburst systems indeed do favor a high incidence of mergers and interactions, morphological signatures of external dynamical perturbations are not detected universally among starbursts (see, e.g., Knapen \& Cisternas 2015; Cibinel \etal 2019; Shangguan \etal 2019).

Lacking clear evidence for an external agent, we turn to the AGN itself as a potential internal trigger of star formation. We anticipate, at the outset, that such an exercise may be frustrated by systematic uncertainties (Harrison \etal 2017) and artificially strong correlations driven by the mutual dependence of SFR and $L_{\rm bol}$ on redshift, stellar mass, and gas content (Zhuang \& Ho 2020; Zhuang et al. 2021).  The wide array of often conflicting results in the literature range from reports of a positive correlation between AGN and star formation activity (\eg Xu \etal 2015; Husemann \etal 2017; Lanzuisi \etal 2017; Schulze \etal 2019; Shangguan \etal 2020b; Stemo \etal 2020; Zhuang \& Ho 2020; Zhuang \etal 2021), to weak or no correlation (\eg Silvermann \etal 2009; Harrison \etal 2012; Mullaney \etal 2012; Rosario \etal 2013a, 2013b; Shimizu \etal 2015; Stanley \etal 2015, 2017), or, indeed, to a negative correlation (\eg Page \etal 2012).  Figure~\ref{fig:ssfr_tdep_agn_par} examines the possible connection between gas depletion time scale and specific SFR on the one hand, and AGN luminosity and Eddington ratio on the other.  No statistically significant correlation between $\tau_{\rm dep}$ and \REdd\ is discerned, but we detect a moderate inverse correlation between $\tau_{\rm dep}$ and $L_{\rm bol}$ ($\tau = -0.26$, $p = 0.001$; Figure~\ref{fig:ssfr_tdep_agn_par}a), consistent with the relation between SFE and BH accretion rate found by Zhuang et al. (2021; their Figure~5). These authors interpret this trend, which in their study is based on a much larger sample of $z \approx 0.3$ type~1 AGNs, as evidence of positive AGN feedback. Our own result is much less secure, as the moderate trend in Figure~\ref{fig:ssfr_tdep_agn_par}a is mainly driven by a handful of points with $L_{\rm bol} \gtrsim  10^{46.3}\ \rm erg\,s^{-1}$.  There seems to be a mildly significant positive correlation between sSFR and $L_{\rm bol}$ and \REdd; the Kendall's test yields a correlation coefficient of $\tau = 0.20$ and $p = 0.05$ for Figure~\ref{fig:ssfr_tdep_agn_par}c and $\tau = 0.23$ and $p = 0.03$ for Figure~\ref{fig:ssfr_tdep_agn_par}d.

In summary, we regard the evidence for a causal connection between star formation and AGN activity to be tenuous, at least insofar as the PG quasar sample itself is concerned.  What seems abundantly clear, from this and many other recent studies, is that in the nearby universe galaxies hosting actively accreting BHs neither lack the fuel nor the ability to form stars.

\section{Conclusions\label{sec:conclusion}}

The SFR and SFE are essential parameters to study how supermassive BHs coevolve with their host galaxies.  We use the sample of 86 $z < 0.5$ PG quasars to investigate the reliability of SFRs based on the total IR (8--1000~$\mum$) emission, through detailed comparison with independent SFRs derive from the [Ne~II]~12.81~$\mum$ and [Ne~III]~15.56~$\mum$ lines following the recent calibration of Zhuang et al. (2019).  We compare the SFRs with the stellar masses and gas masses of the hosts to investigate the nature of their star formation activity in relation to the general galaxy population.  We summarize our main conclusions as follows:

\begin{itemize}
\item The torus-subtracted total IR (8--1000~$\mum$) luminosity yields reliable SFRs for the host galaxies of quasars, at least those with $L_{\rm bol} \approx 10^{44.5} - 10^{47.5}\rm erg\,s^{-1}$.

\item PG quasars form stars at a rate of $\sim 1-250\,M_\odot\,{\rm yr^{-1}}$.  The majority ($75\%-90\%$) of the host galaxies lie on or above the main sequence of local star-forming galaxies, with $\sim 25\%$ having specific SFRs overlapping with those of low-redshift IR-luminous galaxies.

\item In conjunction with their short gas depletion time scales, we estimate that $\sim 75\%$ of the quasar hosts can be considered starburst systems.

\item The stellar morphologies of a significant fraction of the starburst hosts lack evidence of disturbances from recent mergers or interactions.

\item The star formation properties of the present sample of quasars are not strongly coupled to the properties of the AGN.
\end{itemize}

\acknowledgments
We thank our referee for helpful comments. This work was supported by the National Science Foundation of China (11721303, 11991052), the National Key R\&D Program of China (2016YFA0400702), and the National Natural Science Foundation of China for Youth Scientist Project (11803001).  Y.X. thanks Y. Zhao for providing HST morphological types and Tiago Costa, Yingjie Peng, Hassen Yesuf,  Nadia Zakamska, and Chengpeng Zhang for helpful discussions. The Combined Atlas of Sources with Spitzer IRS Spectra (CASSIS) is a product of the IRS instrument team, supported by NASA and JPL. CASSIS is supported by the ``Programme National de Physique Stellaire" (PNPS) of CNRS/INSU co-funded by CEA and CNES and through the ``Programme National Physique et Chimie du Milieu Interstellaire" (PCMI) of CNRS/INSU with INC/INP co-funded by CEA and CNES.

%%% Tables %%%%%%%%%%%%%%%%%%%%%%%%%%%%%%%%%%%%%%%%%%%%%%%%

\input{table_one_input.tex}

\clearpage

\input{table_two_input.tex}

\clearpage

\end{document}

%% file: table_one_input.tex
\begin{deluxetable}{l c r c c l c r c c }
\tablecaption{Physical Properties of PG Quasars \label{tab:basic}}
\tabletypesize{\scriptsize}
\tablehead{
\colhead{Object} &
\colhead{$z$} &
\colhead{$D_L$} &
\colhead{log $\lambda L_\lambda$(5100 \AA)} &
\colhead{log $M_\mathrm{BH}$} &
\colhead{log $M_*$} &
\colhead{12+log (O/H)} &
\colhead{log $M_\mathrm{gas}$} &
\colhead{Morphology} &
\colhead{Reference} \\
\colhead{} &
\colhead{} &
\colhead{(Mpc)} &
\colhead{(erg s$^{-1}$)} &
\colhead{($M_\odot$)} &
\colhead{($M_\odot$)} &
\colhead{} &
\colhead{($M_\odot$)} &
\colhead{} &
\colhead{} \\ 
\colhead{(1)} &
\colhead{(2)} &
\colhead{(3)} &
\colhead{(4)} &
\colhead{(5)} &
\colhead{(6)} &
\colhead{(7)} &
\colhead{(8)} &
\colhead{(9)} &
\colhead{(10)} 
}
\startdata
PG~0003$+$158 & 0.450 & 2572 &        45.99 &       9.45 &  11.83$^{\ast}$   & $ 8.90_{-0.13}^{+0.13} $   &              $<$11.0         &   \nodata    &      \nodata    \\ %S & 
PG~0003$+$199 & 0.025 &  113 &        44.17 &       7.52 &  10.38$^{\ast}$   & $ 8.81_{-0.14}^{+0.14} $   &    $  8.32_{-0.20 }^{+0.20}$ &         D    &           3     \\ %Q & 
PG~0007$+$106 & 0.089 &  420 &        44.79 &       8.87 &           11.03   & $ 8.90_{-0.13}^{+0.13} $   &    $  9.76_{-0.22 }^{+0.22}$ &         M    &           1     \\ %F & 
PG~0026$+$129 & 0.142 &  693 &        45.07 &       8.12 &           11.07   & $ 8.90_{-0.13}^{+0.13} $   &    $  8.90_{-0.22 }^{+0.22}$ &         E    &           3     \\ %Q & 
PG~0043$+$039 & 0.384 & 2133 &        45.51 &       9.28 &           11.13   & $ 8.90_{-0.13}^{+0.13} $   &              $<$10.7         &         U    &           6     \\ %Q & 
PG~0049$+$171 & 0.064 &  297 &        43.97 &       8.45 &  11.07$^{\ast}$   & $ 8.90_{-0.13}^{+0.13} $   &    $  8.66_{-0.36 }^{+0.36}$ &   \nodata    &      \nodata    \\ %Q & 
PG~0050$+$124 & 0.061 &  282 &        44.76 &       7.57 &           11.31   & $ 8.90_{-0.13}^{+0.13} $   &    $ 10.30_{-0.20 }^{+0.20}$ &         D    &           5     \\ %Q & 
PG~0052$+$251 & 0.155 &  763 &        45.00 &       8.99 &           11.24   & $ 8.90_{-0.13}^{+0.13} $   &    $ 10.29_{-0.20 }^{+0.20}$ &         D    &           5     \\ %Q & 
PG~0157$+$001 & 0.164 &  811 &        44.95 &       8.31 &           11.72   & $ 8.90_{-0.13}^{+0.13} $   &    $ 10.80_{-0.20 }^{+0.20}$ &         M    &           4     \\ %Q & 
PG~0804$+$761 & 0.100 &  475 &        45.03 &       8.55 &           10.83   & $ 8.89_{-0.13}^{+0.13} $   &    $  8.79_{-0.21 }^{+0.21}$ &         D    &           5     \\ %Q & 
PG~0838$+$770 & 0.131 &  635 &        44.70 &       8.29 &           11.33   & $ 8.90_{-0.13}^{+0.13} $   &    $ 10.21_{-0.20 }^{+0.20}$ &         D    &           6     \\ %Q & 
PG~0844$+$349 & 0.064 &  297 &        44.46 &       8.03 &           10.88   & $ 8.89_{-0.13}^{+0.13} $   &    $ 10.01_{-0.21 }^{+0.21}$ &         M    &           3     \\ %Q & 
PG~0921$+$525 & 0.035 &  159 &        43.60 &       7.45 &  10.32$^{\ast}$   & $ 8.80_{-0.14}^{+0.14} $   &    $  8.85_{-0.21 }^{+0.21}$ &         D    &           3     \\ %Q & 
PG~0923$+$201 & 0.190 &  955 &        45.01 &       9.33 &           11.28   & $ 8.90_{-0.13}^{+0.13} $   &    $  9.00_{-0.22 }^{+0.22}$ &       E,c    &           3     \\ %Q & 
PG~0923$+$129 & 0.029 &  131 &        43.83 &       7.52 &  10.38$^{\ast}$   & $ 8.81_{-0.14}^{+0.14} $   &    $  9.48_{-0.20 }^{+0.20}$ &         D    &           5     \\ %Q & 
PG~0934$+$013 & 0.050 &  229 &        43.85 &       7.15 &  10.10$^{\ast}$   & $ 8.77_{-0.14}^{+0.14} $   &    $  9.48_{-0.20 }^{+0.20}$ &         D    &           5     \\ %Q & 
PG~0947$+$396 & 0.206 & 1045 &        44.78 &       8.81 &           10.92   & $ 8.89_{-0.13}^{+0.13} $   &    $  9.81_{-0.28 }^{+0.28}$ &         U    &           6     \\ %Q & 
PG~0953$+$414 & 0.239 & 1235 &        45.35 &       8.74 &           11.35   & $ 8.90_{-0.13}^{+0.13} $   &    $ 10.05_{-0.35 }^{+0.35}$ &         U    &           5     \\ %Q & 
\enddata
\tablecomments{
Col. (1): object name.
Col. (2): redshift.
Col. (3): luminosity distance.
Col. (4): monochromatic luminosity of the AGN at 5100 \AA, from Shangguan et al. (2018).
Col. (5): mass of the BH, from Shangguan et al. (2018).
Col. (6): stellar mass of the quasar host galaxy from Zhang \etal (2016) or derived from the $M_{\ast} - M_{\mathrm{BH}}$ relation of Greene et al. (2020; marked with an asterisk).
Col. (7): metallicity of the quasar host galaxy estimated from the stellar mass-metallicity relation; see Section~2 for details. 
Col. (8): total gas mass, from Shangguan et al. (2018).
Col. (9): morphological type of the host galaxy:  ``E" = elliptical, ``D" = disk,  ``M" = merger, ``U" = uncertain,  ``c" = companion, and  ``t" = tidal disturbance signatures. 
Col. (10): reference for morphological types: (1) Bentz \& Manne-Nicholas 2018; (2) Crenshaw et al. 2003; (3) Kim et al. 2017; (4) Surace et al. 1998; (5) Y. Zhao et al. 2021; and (6) Zhang et al. 2016. (Table~1 is published in its entirety in machine-readable format. A portion is shown here for guidance regarding its form and content.) 
}
\end{deluxetable}

%% file: table_two_input.tex
\begin{deluxetable}{l c c c c c c c r}%l }
\tablecaption{SFRs of Quasar Host Galaxies  \label{tab:sfrqso}}
\tabletypesize{\scriptsize}
\tablehead{
\colhead{Object} &
\colhead{log $ L_{\rm [Ne~II]}$}  &
\colhead{log $ L_{\rm [Ne~III]}$}  &
\colhead{log $ L_{\rm [Ne~V]}$ }  &
\colhead{$ f_{\rm +}$}  &
\colhead{$ f_{\rm +2}$} &
\colhead{log\ SFR$_{\rm Ne}$} &
\colhead{log $ L_{\rm IR}$}  &
\colhead{log\ SFR$_{\rm IR}$} \\
\colhead{} &
\colhead{(erg s$^{-1}$)} &
\colhead{(erg s$^{-1}$)} &
\colhead{(erg s$^{-1}$)} &
\colhead{} &
\colhead{} &
\colhead{($M_{\odot}\ \rm yr^{-1}$)} & 
\colhead{(erg s$^{-1}$)} &
\colhead{($M_{\odot}\ \rm yr^{-1}$)} \\
\colhead{(1)} &
\colhead{(2)} &
\colhead{(3)} &
\colhead{(4)} &
\colhead{(5)} &
\colhead{(6)} &
\colhead{(7)} & 
\colhead{(8)} & 
\colhead{(9)}
}
\startdata
 PG~0003$+$158        &                           \nodata  &                            \nodata  &                            \nodata  &                           \nodata  &                            \nodata  &                                      \nodata  &                         $< 44.60 $  &                          $< 1.25 $ \\
 PG~0003$+$199        &         $ 40.59_{-0.10}^{+0.10} $  &          $ 40.82_{-0.03}^{+0.03} $  &          $ 40.48_{-0.07}^{+0.07} $  &          $ 0.46_{-0.08}^{+0.08} $  &           $ 0.47_{-0.09}^{+0.09} $  &                     $ 0.29_{-0.17}^{+0.17} $  &          $ 43.11_{-0.03}^{+0.03} $  &          $ -0.24_{-0.03}^{+0.03} $ \\
 PG~0007$+$106        &         $ 41.53_{-0.06}^{+0.06} $  &          $ 41.91_{-0.03}^{+0.03} $  &          $ 41.55_{-0.05}^{+0.05} $  &          $ 0.35_{-0.05}^{+0.05} $  &           $ 0.60_{-0.06}^{+0.06} $  &                     $ 1.19_{-0.15}^{+0.15} $  &          $ 44.27_{-0.03}^{+0.03} $  &           $ 0.92_{-0.03}^{+0.03} $ \\
 PG~0026$+$129        &         $ 41.41_{-0.20}^{+0.20} $  &         $ 41.79_{-0.13}^{+0.13} $   &          $ 41.49_{-0.15}^{+0.15} $  &          $ 0.37_{-0.17}^{+0.17} $  &           $ 0.58_{-0.19}^{+0.19} $  &                     $ 1.07_{-0.24}^{+0.24} $   &            $ 43.86_{-0.04}^{+0.03} $  &           $ 0.51_{-0.04}^{+0.03} $ \\ 
 PG~0043$+$039        &                           \nodata  &                            \nodata  &                            \nodata  &                           \nodata  &                            \nodata  &                                      \nodata  &                         $< 44.80 $  &                          $< 1.45 $ \\
 PG~0049$+$171        &                           \nodata  &                            \nodata  &                            \nodata  &                           \nodata  &                            \nodata  &                                      \nodata  &          $ 42.91_{-0.07}^{+0.05} $  &          $ -0.43_{-0.07}^{+0.05} $ \\
 PG~0050$+$124        &                        $< 40.91 $  &          $ 41.35_{-0.14}^{+0.14} $  &                         $< 41.11 $  &                           \nodata  &                            \nodata  &                                  [0.19, 0.92]  &          $ 44.94_{-0.01}^{+0.01} $  &           $ 1.60_{-0.01}^{+0.01} $ \\
 PG~0052$+$251        &                           \nodata  &                            \nodata  &                            \nodata  &                           \nodata  &                            \nodata  &                                      \nodata  &          $ 44.51_{-0.02}^{+0.02} $  &           $ 1.16_{-0.02}^{+0.02} $ \\
 PG~0157$+$001        &         $ 42.81_{-0.03}^{+0.03} $  &          $ 42.92_{-0.03}^{+0.03} $  &          $ 42.62_{-0.06}^{+0.06} $  &          $ 0.55_{-0.03}^{+0.03} $  &           $ 0.38_{-0.03}^{+0.03} $  &                     $ 2.38_{-0.14}^{+0.14} $  &          $ 45.85_{-0.05}^{+0.04} $  &           $ 2.50_{-0.05}^{+0.04} $ \\
 PG~0804$+$761        &                        $< 41.23 $  &          $ 41.69_{-0.09}^{+0.09} $  &                         $< 41.13 $  &                           \nodata  &                            \nodata  &                                  [0.76, 1.25]  &          $ 43.83_{-0.05}^{+0.07} $  &           $ 0.48_{-0.05}^{+0.07} $ \\
 PG~0838$+$770        &         $ 41.10_{-0.15}^{+0.15} $  &          $ 41.37_{-0.15}^{+0.15} $  &          $ 41.24_{-0.13}^{+0.13} $  &          $ 0.43_{-0.20}^{+0.20} $  &           $ 0.51_{-0.21}^{+0.21} $  &                     $ 0.63_{-0.26}^{+0.26} $  &          $ 44.72_{-0.04}^{+0.03} $  &           $ 1.37_{-0.04}^{+0.03} $ \\
 PG~0844$+$349        &         $ 40.37_{-0.25}^{+0.25} $  &          $ 41.09_{-0.10}^{+0.10} $  &          $ 40.68_{-0.17}^{+0.17} $  &          $ 0.18_{-0.12}^{+0.12} $  &           $ 0.80_{-0.14}^{+0.14} $  &                     $ 0.29_{-0.21}^{+0.21} $  &          $ 43.61_{-0.02}^{+0.02} $  &           $ 0.26_{-0.02}^{+0.02} $ \\
 PG~0921$+$525        &         $ 40.85_{-0.04}^{+0.04} $  &          $ 40.94_{-0.04}^{+0.04} $  &          $ 40.43_{-0.08}^{+0.08} $  &          $ 0.54_{-0.04}^{+0.04} $  &           $ 0.39_{-0.04}^{+0.04} $  &                     $ 0.57_{-0.15}^{+0.15} $  &          $ 43.09_{-0.02}^{+0.02} $  &          $ -0.26_{-0.02}^{+0.02} $ \\
 PG~0923$+$201        &                        $< 41.23 $  &          $ 41.70_{-0.19}^{+0.19} $  &                         $< 41.37 $  &                           \nodata  &                            \nodata  &                                  [0.62, 1.25]  &          $ 43.99_{-0.05}^{+0.05} $  &           $ 0.64_{-0.05}^{+0.05} $ \\
 PG~0923$+$129        &         $ 40.96_{-0.03}^{+0.03} $  &          $ 41.03_{-0.03}^{+0.03} $  &          $ 40.84_{-0.03}^{+0.03} $  &          $ 0.58_{-0.03}^{+0.03} $  &           $ 0.35_{-0.03}^{+0.03} $  &                     $ 0.56_{-0.15}^{+0.15} $  &          $ 44.05_{-0.02}^{+0.01} $  &           $ 0.70_{-0.02}^{+0.01} $ \\
 PG~0934$+$013        &                           \nodata  &                            \nodata  &                            \nodata  &                           \nodata  &                            \nodata  &                                      \nodata  &          $ 43.96_{-0.02}^{+0.02} $  &           $ 0.61_{-0.02}^{+0.02} $ \\
 PG~0947$+$396        &                           \nodata  &                            \nodata  &                            \nodata  &                           \nodata  &                            \nodata  &                                      \nodata  &          $ 44.29_{-0.04}^{+0.05} $  &           $ 0.94_{-0.04}^{+0.05} $ \\
 PG~0953$+$414        &         $ 41.45_{-0.29}^{+0.29} $  &          $ 42.00_{-0.12}^{+0.12} $  &          $ 41.55_{-0.30}^{+0.30} $  &          $ 0.30_{-0.15}^{+0.15} $  &           $ 0.65_{-0.17}^{+0.17} $  &                     $ 1.29_{-0.24}^{+0.24} $  &          $ 44.30_{-0.07}^{+0.07} $  &           $ 0.95_{-0.07}^{+0.07} $ \\
\enddata
\tablecomments{
Col. (1): object name.
Col. (2): luminosity of [Ne~II]\,12.81~$\mum$.
Col. (3): luminosity of [Ne~III]\,15.56~$\mum$.
Col. (4): luminosity of [Ne~V]\,14.32~$\mum$.
Col. (5): fractional abundance of neon in the Ne$^{+}$ state.
Col. (6): fractional abundance of neon in the Ne$^{+2}$ state.
Col. (7): SFR calculated from neon luminosity. For SFR involving neon line upper limits, the values included in brackets present lower and upper limits of the true SFR. 
Col. (8): torus-subtracted IR luminosity in the $8-1000\,\mum$ band.
Col. (9): SFR calculated from the torus-subtracted IR luminosity in the $8-1000\,\mum$ band. 
(Table~2 is published in its entirety in machine-readable format. A portion is shown here for guidance regarding its form and content.)
}
\end{deluxetable}